\documentclass[preprint,showpacs,preprintnumbers,
nofootinbib, 
amsmath,amssymb,color]{revtex4-1}

\usepackage{graphicx}
\usepackage{dcolumn}
\usepackage{bm,morefloats,color}

\usepackage{amssymb,bm,color}

\def\eqref#1{Eq.~(\ref{#1})}

\def\Eq#1{\begin{equation} #1 \end{equation}}
\def\Eqr#1{\begin{eqnarray} #1 \end{eqnarray}}
\def\Eqrsubl#1#2{\begin{subequations}\label{#1}\Eqr{#2}\end{subequations}}

\newcommand{\nn}{\nonumber}
\newcommand{\pd}{\partial}

\def\X5sp{{\rm X}_5}
\def\Y3sp{{\rm Y}_3}
\def\Z3sp{{\rm Z}_3}

\def\lap{{\triangle}}
\def\e{{\rm e}}

\begin{document}


\title{
Dynamical 
de Sitter conjecture and 
quintessence model
}

\author{Muneto Nitta$^{a,b,c}$}%
\author{Kunihito Uzawa$^{d,b}$}
\affiliation{%
$^a$Department of Physics, and Research and Education 
Center for Natural Sciences, Keio University, 
Hiyoshi 4-1-1, Yokohama, Kanagawa 223-8521, Japan}%
\affiliation{%
$^b$Research and Education 
Center for Natural Sciences, Keio University, 
Hiyoshi 4-1-1, Yokohama, Kanagawa 223-8521, Japan}%
\affiliation{
$^c$ International Institute for Sustainability with Knotted Chiral Meta Matter(WPI-SKCM$^2$), Hiroshima University, 1-3-2 Kagamiyama, Higashi-Hiroshima, Hiroshima 739-8511, Japan
}

\affiliation{%
$^d$ Department of Physics,
School of Science and Technology,
Kwansei Gakuin University, Sanda, Hyogo 669-1337, Japan}

\date{\today}

\begin{abstract}
The de Sitter conjecture yields a severe bound on possible vacua for 
a consistent quantum gravity.
We extend 
the de Sitter conjecture by taking into account 
dynamics of the scalar field.
We then apply such an extended de Sitter conjecture 
to a quintessence model of inflation for which dynamics of 
the scalar field is essential, 
and obtain an allowed region of parameters 
of the scalar potential wider than previously considered cases 
with the conventional de Sitter conjecture.
The new bounds in the 
swampland conjecture could have implications in several 
situations to construct compactification models.

\end{abstract}

\maketitle


\section{Introduction}
 \label{sec:introduction}
String theory is one of the candidates for quantum field theories consistent with theory of gravity. Since it is defined in higher-dimensional spacetimes, a compactification process becomes necessary to ensure compatibility with our four-dimensional 
universe. Each compactification represents a different low-energy effective field theory in four dimensions, which generates what is called the ``landscape'' of string theory \cite{Susskind:2003kw, Banks:2003es, Denef:2004ze, Kallosh:2004yh, Vafa:2005ui, Denef:2008wq, Brennan:2017rbf, vanBeest:2021lhn, Grana:2021zvf, Agmon:2022thq}. Although the four-dimensional quantum field theory describing the standard model of particle physics has succeeded in explaining various interactions, it is also true that numerous problems remain in field theories beyond the standard model, such as the need for parameter fine-tuning to determine coupling constants and particle masses.

%
%
In recent developments of string theory, it has been suggested that not all quantum field theories are necessarily consistent with gravitational theories, and there exist conditions for four-dimensional field theories to be gravitationally consistent. Field theories that do not satisfy these conditions are referred to as residing in the ``swampland'' \cite{Vafa:2005ui, Ooguri:2006in, Ooguri:2016pdq, Brennan:2017rbf, Obied:2018sgi, Agrawal:2018own, Ooguri:2018wrx, Andriot:2018wzk, Palti:2019pca, Bedroya:2019snp, Agmon:2022thq}. Some solutions to higher-dimensional string theories become theories within the landscape through compactification, while others become theories residing outside the landscape, i.e., in the swampland. 
The latter 
are no longer consistent with gravitational theories. The classification of quantum field theories consistent with quantum gravity potentially provides powerful theoretical tools, offering strategies to resolve fine-tuning problems \cite{Heckman:2019bzm} and enabling predictions about cosmological observations by clarifying the swampland region, such as the dynamics of early universe inflation and the nature of dark energy \cite{Agrawal:2019dlm, Bedroya:2019snp, Vafa:2019evj, McNamara:2020uza, vanBeest:2021lhn, Agmon:2022thq}.
%
%
Although string theory might provide the field theory of the standard model of particle physics, no one knows the full extent of the landscape. If the landscape were to encompass all quantum field theories, string theory would find it challenging to provide insights into field theories beyond the standard model. However, we cannot obtain the low-energy limits of all string-theory compactifications, and it is becoming increasingly clear that numerous field theories exist outside the landscape \cite{Vafa:2005ui, Palti:2019pca}.
%
%

Consequently, the most critical issue in current string theory is determining the boundary between the landscape and the swampland. However, given the countless possible compactifications of string theory, definitively establishing this boundary is not straightforward \cite{Brennan:2017rbf}. Therefore, there are systematic investigations of reliable compactifications that provide realistic four-dimensional universe models and standard model consistent with observations and experiments, empirically confirming the characteristics of four-dimensional field theories they share. From these data, studies are developing to hypothesize criteria for determining which field theories belong to the landscape or swampland and to define the boundary between them \cite{Brennan:2017rbf, Palti:2019pca, Bedroya:2019snp}.
%
%

As an example of determining such a boundary, 
the so-called de Sitter conjecture has been proposed.
This conjecture addresses the possibility that the de Sitter vacuum becomes part of the swampland in the quantum field theory under consideration, and it is derived from the field equations and energy conditions of string theory \cite{Obied:2018sgi}. The de Sitter conjecture is further supported by discussions of distance conjectures related to the de Sitter spacetime entropy and the dynamics of extra dimensions \cite{Ooguri:2018wrx}. For instance, as extra dimensions increase in size, light particles emerge on the four-dimensional spacetime \cite{Ooguri:2006in}, and their contributions impact the dynamics of early universe inflation. These new particles are colloquially referred to as the ``tower of light states'' \cite{Ooguri:2006in} and play a role in constraining inflationary models \cite{Agrawal:2018own, Garg:2018reu}.
%
%
The de Sitter conjecture provides constraints on the low-energy effective theories of consistent quantum gravitational theories, and these constraints (that is, the criteria for determining whether a theory belongs to the landscape or swampland) stem from the current accelerating expansion of the universe. In other words, the low-energy effective theories derived from consistent quantum gravity theories are evaluated on the basis of whether they can reproduce the observed accelerating universe and the inflation of the early universe without contradiction. The criteria determine whether a quantum field theory resides in the landscape or in the swampland. Naturally, if a low-energy effective theory realizes a meta-stable de Sitter vacuum \cite{Kachru:2003aw, Kachru:2003sx} and accelerated expansion at the present universe, it would be considered part of the landscape. Conversely, if it fails to do so, the four-dimensional quantum field theory would be deemed to reside in the swampland \cite{Obied:2018sgi, Andriot:2018wzk, Garg:2018reu, Palti:2019pca}.

The de Sitter conjecture gives a lower bound on the gradient of potentials in positive regions 
\cite{Obied:2018sgi}. The scalar potential of a theory coupled to gravity have to satisfy a
bound on its derivatives with respect to scalar fields
\Eq{
|\nabla V|>c\,V\,,
  \label{swampland}
}
where $c>0$ is of order 1 in Planck units. The number $c$ generally depends on the number of dimensions of background space-time and internal space. This bound on $\left|\nabla V\right|/V$ for 
compactifications of string theory is in fact a consequence of strong energy condition 
(SEC) in higher dimensions and field equations. The criteria can be derived by a simple generalization of Maldacena \& Nu\~nez \cite{Maldacena:2000mw} and Steinhardt-Wesley 
\cite{Steinhardt:2008nk} no-go theorems that one can place a bound on the gradient of the potential. 

As mentioned above, the de Sitter conjecture is fundamentally constructed based on the field equations of the original higher-dimensional theory, the low-energy effective theory obtained after compactification, and the 
SEC \cite{Obied:2018sgi}. In this process, the time derivative of the scalar field is set to zero. The scalar field represents the moduli in the higher-dimensional theory, corresponding to the scale of the internal space, is fixed \cite{Obied:2018sgi}. 
If the scalar field coming from moduli triggers an inflaton, the condition of dropping off the time derivative would be consistent with the slow-roll condition of inflation. 

The early universe inflationary expansion are achieved by the dynamics of scalar fields, particularly through scalar fields which possess minimum or local minimum in their potential \cite{Kachru:2003aw, Kachru:2003sx}. The potential energy at these minima or local minima of scalar fields is equivalent to a positive cosmological constant in four-dimensional effective theories, thus leading to stable or meta-stable de Sitter spacetimes. In string theory, there are numerous efforts to make stable or meta-stable de Sitter vacua \cite{Kachru:2003aw, Kachru:2003sx}, but an overwhelming number of problems still have emerged. Within the issues of reproducing stable or meta-stable de Sitter vacua in string theory, recent proposals have focused on constraints imposed on scalar field potentials to avoid residing in the swampland. 

On the other hand, 
the current accelerating cosmic expansion should be explained by a potential of the scalar field that lacks minimum or local minimum; If the potential remains consistently positive and the contribution from its kinetic term is not overly significant, one can still explain an accelerating universe consistent with current observational data. 
Such scalar fields are often referred to as quintessence 
\cite{Caldwell:1997ii, Carroll:1998zi, Zlatev:1998tr, Chiba:1999ka, Tsujikawa:2013fta}.

The simplest and most natural choice of the potential for quintessence models is 
given by 
an exponential function \cite{Barreiro:1999zs, Raveri:2018ddi}
\Eqr{
V=V_0\,\exp\left(-\lambda \tau\right)\,,
  \label{qp}
}
where  $\tau$ is a scalar field  
and $\lambda$ is a constant slope. 
However, a puzzle issue has 
recently emerged with the 
quintessence models with 
the above scalar potential.  
The allowed region 
$\lambda < 0.60$
of the potential slope $\lambda$ 
in the numerical solutions of the field equations in four-dimensional dynamical systems
\cite{Andriot:2024jsh, Rudelius:2022gbz, Shiu:2023nph, Copeland:1997et, Ferreira:1997hj, Bahamonde:2017ize, Alestas:2024gxe, Casas:2024oak} suggested by observational facts of supernovae \cite{DES:2024jxu} and BAO \cite{DESI:2024mwx} 
is not consistent with the result 
$\lambda \geq \sqrt 2$  
constrained by  
{\bf the trans-Planckian censorship conjecture and 
the strong de Sitter conjecture 
\cite{Bedroya:2019snp, Rudelius:2021oaz}}. 
However, when considering quintessence, vanishing the time derivative of the scalar field is not necessarily a condition that must be satisfied. Here there is also a problem with quintessence as dark energy/late time accelerated expansion is that it is guaranteed to make $H_0$ tension. These issues are explored in
\cite{Colgain:2019joh, Banerjee:2020xcn, Lee:2022cyh}. They slowly generalized the statement from the potential (\ref{qp}), to generic potentials, and beyond quintessence to the Horndeski class of models.  

This note studies the de Sitter conjecture with dynamical scalar field,
when removing the assumption of neglecting the time derivative 
on it. 
When the time derivative of scalar field is non-zero, the energy conditions and field theory equations of higher-dimensional and the low energy effective theories after compactification cannot neglect the kinetic term of scalar field, and these contributions 
give a correction to
the de Sitter conjecture 
in Eq.~(\ref{swampland}). 
We find that if we apply such ``dynamical de Sitter conjecture'' to the quintessence model in which the time derivative is not negligible, 
we obtain an allowed region of the parameters broader 
than previously considered 
($\lambda \geq \sqrt 2$). 
Our results provide a solution to 
the aforementioned discrepancy 
between observations and 
the de Sitter conjecture, owing to expanding the allowed region of the parameter space for the scalar field potential. 

In section \ref{sec:sw}, we revisit the original de Sitter conjecture, explicitly presenting the equations of motion and strong energy conditions without neglecting the time derivative terms of the scalar field. We discuss how the kinetic terms of the scalar field contribute to the criteria of de Sitter conjecture. Section \ref{sec:quintessence} addresses the quintessence model when the potential of the scalar field is expressed exponentially and determines the allowed region of potential slopes when applying the dynamical de Sitter conjecture to this model. In Section \ref{sec:summary}, we give cosmological considerations and future implications based on our discussions.

\section{de Sitter conjecture and Extension of Swampland criteria}
\label{sec:sw}

In this section, we discuss the de Sitter conjecture or swampland criteria.
After reviewing it in the first subsection, 
we extend it in the case of scalar field is dynamical in the second subsection.


Now we suppose a theory in $D$ dimensions and compactify 
it on a $(D-d)$-dimensional space with a metric $g_{mn}$ and 
a warp factor $\Omega$ as,
\Eq{
ds^2=g_{MN}dx^Mdx^N=
\Omega^2(t, y)
\left[-dt^2+a^2(t)\delta_{\alpha\beta}dx^\alpha dx^\beta\right]
+g_{mn}(t, y)dy^mdy^n\,,
 \label{se:metric:Eq}
}
where $g_{mn}$ denotes the metric of internal space, 
$x^M$\,, $x^{\alpha}$\, and $y^m$\, are the coordinates of 
$D$-dimensional spacetime, $(d-1)$-dimensional space, 
$(D-d)$-dimensional internal space, respectively. 
We assume that the warp factor $\Omega$ and the internal space metric $g_{mn}$ depend only on the
coordinates of the internal space $y^m$ and time $t$,.

We fix a residual diffeomorphism by choosing the 
$d$-dimensional gravitational constant $\kappa_d$ to 
be equal to $D$-dimensional one $\kappa_D$. This implies 
\cite{Obied:2018sgi}
\Eq{
\int d^{D-d}y\,\Omega^{d-2}\sqrt{{\rm det}\,g_{mn}}=1\,.
   \label{se:gc:Eq}
}
Hence we have 
\Eqr{
&&\frac{\pd^2}{\pd t^2}\left(\int d^{D-d}y \Omega^{d-2}\sqrt{\det g_{mn}}
\right)=\int d^{D-d}y \Omega^{d-2}\sqrt{\det g_{mn}}\left[
\frac{1}{2}\left(g^{mn}\ddot{g}_{mn}+\dot{g}^{mn}\dot{g}_{mn}\right)
\right.\nn\\
&&\left.~~~~~~+\frac{1}{4}\left(g^{mn}\dot{g}_{mn}\right)^2
+(d-2)\Omega^{-1}\ddot{\Omega}+(d-3)\Omega^{-2}\dot{\Omega}^2
\right]=0\,,
    \label{se:residual:Eq}
}
where we will assume time dependences 
\Eq{
\dot{\Omega}\equiv \pd_t\Omega \ne 0\,,~~~~~
\dot{g}_{mn}\equiv\pd_t g_{mn}\ne 0\,.
  \label{se:condition:Eq}
}

The computation of the Ricci tensor $R_{tt}$ corresponding 
the background metric (\ref{se:metric:Eq}) gives
\Eqr{
R_{tt}&=&-\left(d-1\right)\left[
a^{-1}\ddot{a}+\left(a\,\Omega\right)^{-1}\dot{a}\,\dot{\Omega}
-\Omega^{-2}\,\dot{\Omega}^2+\Omega^{-1}\,\ddot{\Omega}\right]\nn\\
&&-\frac{1}{2}\left(g^{mn}\ddot{g}_{mn}+\dot{g}^{mn}\dot{g}_{mn}
-\Omega^{-1}\,g^{mn}\dot{g}_{mn}\,\dot{\Omega}
-\frac{1}{2}g^{mm'}\,g^{nn'}\,\dot{g}_{nm'}\dot{g}_{mn'}\right)\nn\\
&&+\Omega\,\bar{\lap}\Omega+(d-1)g^{mn}\pd_m\Omega\,\pd_n\Omega\,,
}
where $\bar{\lap}$ indicates $(D-d)$-dimensional Laplace operator 
with respect to the metric $g_{mn}$.
We will try to obtain a restriction on the values of the 
acceleration using the strong energy condition in 
$D$-dimensions. It gives
\Eqr{
0\le R_{tt}&=&-\left(d-1\right)\left[
a^{-1}\ddot{a}+\left(a\,\Omega\right)^{-1}\dot{a}\,\dot{\Omega}
-\Omega^{-2}\,\dot{\Omega}^2+\Omega^{-1}\,\ddot{\Omega}\right]\nn\\
&&-\frac{1}{2}\left(g^{mn}\ddot{g}_{mn}+\dot{g}^{mn}\dot{g}_{mn}
-\Omega^{-1}\,g^{mn}\dot{g}_{mn}\,\dot{\Omega}
-\frac{1}{2}g^{mm'}\,g^{nn'}\,\dot{g}_{nm'}\dot{g}_{mn'}\right)\nn\\
&&+\Omega\,\bar{\lap}\Omega+(d-1)g^{mn}\pd_m\Omega\,\pd_n\Omega\,,
  \label{asw:st:Eq}
}
where we have used the $D$-dimensional Einstein equations. 

With multiplying Eq.~(\ref{asw:st:Eq}) by $\Omega^{d-2}$, 
integrating over the compact manifold and using 
a residual diffeomorphism (\ref{se:residual:Eq}), we have
\Eqr{
&&(d-1)a^{-1}\ddot{a}\le \int d^{D-d}y \Omega^{d-2}\sqrt{\det g_{mn}}
\left[\Omega^{-1}\left\{\ddot{\Omega}+(d-1)H\dot{\Omega}\right\}
+2(d-2)\Omega^{-2}\dot{\Omega}^2\right.
\nn\\
&&\left.~~~~~~
+\frac{1}{4}\left(g^{mn}\dot{g}_{mn}\right)^2
+\frac{1}{2}\left(\Omega^{-1}g^{mn}\dot{g}_{mn}\dot{\Omega}
+\frac{1}{2}g^{mm'}g^{nn'}\dot{g}_{nm'}\dot{g}_{mn'}\right)\right],
  \label{asw:st2:Eq}
}
where we have used the fact that the integral of the Laplacian
of the warp factor over the whole manifold is zero, 
because the manifold has no boundary
and the warp factor is non-singular.

We can obtain a lower bound on the second derivative of the 
warp factor which is related to the acceleration of the scalar 
fields by using the $D$-dimensional Einstein equations. 
Here we will discuss the $d$-dimensional effective action 
to relate this restriction to a bound on the potential. 

Since the inequality (\ref{asw:st2:Eq}) gives a bound on the value 
of the scalar field acceleration only in the direction of the 
overall volume change, other deformations of the manifold
are not important. 
We will define a scalar mode $\tau$ as a basis, which is 
the overall volume modulus while other fields are chosen to 
be orthogonal to it. Now it is necessary for us to find the 
$d$-dimensional effective action in order to relate the 
bound on the acceleration to a bound on the 
derivative of the potential. As was noted above, 
only overall scaling modulus $\tau$ will be analyzed. 
Separating the modulus $\tau$ from the others, we find
\Eqr{
ds^2&=&\e^{-2\sqrt{\frac{D-d}{(D-2)(d-2)}}\,\tau}\,
\tilde{\Omega}^2(\Phi, y) \left[-dt^2+a^2(t)
\delta_{\alpha\beta}dx^\alpha dx^\beta \right]\nn\\
&&+\e^{2\sqrt{\frac{d-2}{(D-d)(D-2)}}\,\tau}\,
\tilde{g}_{mn}(\Phi, y)dy^mdy^n\,.
}
Here, the tilde denotes that the overall modulus 
contribution was singled out. 
{\bf In what follows, we assume that the moduli 
$\Phi$ is fixed and focus only on the dynamics of $\tau$.
} The
relevant part of the $d$-dimensional effective action is
\Eq{
S=\int d^dx\sqrt{-{\rm det}\,g_d}\left[
\frac{1}{2\kappa_d^2}R_d+\frac{1}{2\kappa_d^2}
\dot{\tau}^2-V(\tau, \Phi)+\cdots  
\right],
  \label{se:d-action:Eq}
}
where $\kappa_d$ is the $d$-dimensional gravitational 
coupling constant and $\Phi$ denotes for all other fields.
The $d$-dimensional Einstein equations and the equation 
of motion for the scalar field $\tau$ are 
\Eqrsubl{asw:E:Eq}{
&&R_{tt}=-(d-1)a^{-1}\ddot{a}=\dot{\tau}^2
-\frac{2\kappa_d^2}{d-2}V\,,\\
&&R_{ij}=\left[a^{-1}\ddot{a}+(d-2)H^2\right]g_{ij}
=\frac{2\kappa_d^2}{d-2}g_{ij}V\,,\\
&&\ddot{\tau}+(d-1)H\dot{\tau}+\kappa_d^2\pd_\tau V=0\,.
   \label{asw:scalar:Eq}
}
From the Einstein equations (\ref{asw:E:Eq}), 
we have 
\Eq{
H^2=\frac{1}{(d-1)(d-2)}\left(\dot{\tau}^2+2\kappa_d^2V\right)\,,
~~~a^{-1}\ddot{a}=H^2-(d-2)^{-1}\dot{\tau}^2\,.
   \label{asw:Ein:Eq}
}
Then, we obtain 
\Eqr{
a^{-1}\ddot{a}=\frac{1}{(d-1)(d-2)}\left[-(d-2)\dot{\tau}^2
+2\kappa_d^2 V\right].
}
The left hand side of Eq.~(\ref{asw:st2:Eq}) becomes 
\Eq{
(d-1)a^{-1}\ddot{a}=-\dot{\tau}^2+\frac{2\kappa_d^2}{d-2}V\,.
  \label{asw:lhs:Eq}
}
If we use the scalar field equation (\ref{asw:scalar:Eq}) and 
the definition of modulus $\tau$\,, 
we have 
\Eqrsubl{asw:mo:Eq}{
\Omega(t, y)&=&\exp\left[-\sqrt{\frac{D-d}{(D-2)(d-2)}}\,\tau\right]
\tilde{\Omega}(\Phi, y)\,,\\
g_{mn}(t, y)&=&\exp\left[2\sqrt{\frac{d-2}{(D-d)(D-2)}}\,\tau\right]
\tilde{g}_{mn}(\Phi, y)\,,
}
{\bf where 
note that we assume here that the moduli 
$\Phi$ other than $\tau$ are fixed.
}

The right hand side of Eq.~(\ref{asw:st2:Eq}) can be rewritten as 
\Eqr{
\left[\frac{-d^3+d^2(D-1)+6D(d-2)+4}{4(D-2)(d-2)}\right]
\dot{\tau}^2
+\sqrt{\frac{D-d}{(D-2)(d-2)}}\kappa_d^2\pd_\tau V\,,
   \label{asw:rhs:Eq}
}
{\bf Here, we assume that the moduli 
$\Phi$ other than $\tau$ are fixed, and 
for the integral over the extra-dimensional 
space, we have used (\ref{se:gc:Eq}).}

Combining Eqs.~(\ref{asw:lhs:Eq}) and 
(\ref{asw:rhs:Eq}) with the SEC bound in Eq.~(\ref{asw:st2:Eq}), 
we obtain the following bound on the gradient of the potential,
\Eq{
-\frac{\pd_\tau V}{V}\ge 
2\sqrt{\frac{D-2}{(D-d)(d-2)}}
-\alpha\frac{\dot{\tau}^2}{V}\,,
  \label{asw:new:Eq}
}
where $\alpha$ is defined by 
\Eq{
\alpha=\frac{1}{4\kappa_d^{2}}\sqrt{\frac{D-d}{(D-2)(d-2)}}
\left[-5d^2+d\left(5D+12\right)-11D-2\right].
}

If the scalar field $\tau$ satisfies the slow-roll condition 
\Eq{
\frac{\dot{\tau}^2}{V}\ll 1\,,
}
the bound in Eq.~(\ref{asw:new:Eq}) reduces to the well-known swampland criteria 
\cite{Obied:2018sgi}
\Eq{
-\frac{\pd_\tau V}{V}
\ge 2\sqrt{\frac{D-2}{(D-d)(d-2)}}
=\lambda_{\rm SEC}\,.
}
When the velocity of the scalar field is not negligible, 
we have to use \eqref{asw:new:Eq} 
that we call dynamical de Sitter conjecture.


\section{Application to Quintessence}
\label{sec:quintessence} 

In this section, we discuss an application of 
dynamical de Sitter conjecture in 
Eq.~(\ref{asw:new:Eq}) obtained in the last section.

As an example in which dynamics of the scalar field 
is essential, here we focus on the quintessence model for inflation. 
In such a case, 
Eqs.~(\ref{asw:scalar:Eq}) and (\ref{asw:Ein:Eq}) 
can be solved exactly
which the $d$-dimensional exponential potential
\cite{Abbott:1984fp, Lucchin:1984yf, 
Lucchin:1985wy, Lyth:1991bc} 
\Eqr{
V=V_0\,\exp\left(-\lambda \tau\right)\,,
   \label{ap:potential:Eq}
}
with $V_0$ and $\lambda>0$ arbitrary constants. 
It is easy to verify that Eqs.~(\ref{asw:scalar:Eq}) 
and (\ref{asw:Ein:Eq}) are satisfied by
\footnote{{\bf
This is not the only possible solution   
which have the attractive fixed point at the late time. 
Since this is correct for $\varepsilon<d-1$\,, 
we find an upper bound to $\lambda$\,. 
Note that there is a kination solution, as well as solutions 
going from the latter to the former upon setting 
the potential (\ref{ap:potential:Eq}).}
}
\Eqr{
\tau(t)=\frac{1}{\lambda}\ln\left(\tau_0 t^2\right)\,,~~~~~
\tau_0=\frac{2\kappa_d^2\,V_0\,\varepsilon^2}
{(d-2)(d-1-\varepsilon)}\,,
}
and 
\Eqr{
a\propto t^{1/\varepsilon}\,,~~~~~
H=\frac{1}{\varepsilon t}\,,
}
where $\varepsilon$ is the positive dimensionless 
quantity
\Eqr{
\varepsilon\equiv \frac{d-2}{4}\,\lambda^2\,.
}
The extra term $\dot{\tau}^2/V$ becomes 
\Eq{
\frac{\dot{\tau}^2}{V}=\frac{4\tau_0}{\lambda^2\,V_0}
=-\frac{2(d-2)\kappa_d^2\lambda^2}
{4-2\lambda^2+d(\lambda^2-4)}\,.
}
The value of the term $\frac{\dot{\tau}^2}{V}$ for each total dimension $D$ is listed in table~\ref{table_new}\,, where we set $d=4$ and $\kappa_4^2=1$ and $\lambda$ takes the smallest positive value in each dimension. One can see that as $D$ decreases, 
the kinetic term of the scalar field $\frac{\dot{\tau}^2}{V}$ cannot be neglected for any value of $V_0$\,. 
\begin{table}[ht]
\caption{We show the contribution of $\dot{\tau}^2/V$ for $d=4$\,, 
$\kappa_4^2=1$\,, when $\lambda$ is the smallest positive value. 
{\bf Here we present $\dot{\tau}^2/V$ with respect to smallest value 
of $\lambda$ because it takes minimum for $\lambda<2$\,.} 
As the number of total dimension $D$ 
decreases, the term $\dot{\tau}^2/V$ becomes large.}
{\scriptsize
\begin{center}
\begin{tabular}{|c|c|c|c|c|c|c|c|}
\hline
$D$&5&6&7&8&9&10&11
\\
\hline
$\dot{\tau}^2/V$ & 0.945619 & 0.4 & 0.253046 & 0.185636 & 0.147017 & 0.121975 & 0.104402 
\\
\hline
\end{tabular}
\end{center}
}
\label{table_new}
\end{table}

The potential of the scalar field depends on time
\Eqr{
V(t)
=\bar{V_0}\,t^{-2}\,,
}
where $\bar{V_0}$ denotes 
\Eqr{
\bar{V_0}\equiv\frac{V_0}{\tau_0}=
\frac{8\left(d-1-\frac{d-2}{4}\lambda^2\right)}
{(d-2)\kappa_d^2\,\lambda^4}\,.
}
Upon setting 
\Eqr{
b=2\sqrt{\frac{D-2}{(D-d)(d-2)}}\,,
}
we can find that the swampland criteria 
(\ref{asw:new:Eq}) gives
\Eqr{
\lambda\ge b-\frac{2\alpha\,\kappa_d^2\,\lambda^2}
{\frac{4(d-1)}{d-2}-\lambda^2}\,.
}
Then, the slope of the scalar potential $\lambda$
has to obey
\Eqr{
&&\lambda^3-(2\alpha\,\kappa_d^2+b)\lambda^2
-\frac{4(d-1)}{d-2}\kappa_d^2\lambda
+\frac{4(d-1)}{d-2}b\le 0\,.
}

Let us define the function $f(\lambda)$\,, 
\Eqr{
f(\lambda)=\lambda^3-(2\alpha\,\kappa_d^2+b)\lambda^2
-\frac{4(d-1)}{d-2}\kappa_d^2\lambda
+\frac{4(d-1)}{d-2}b\,,
}
we will find the allowed region of $\lambda$ fulfilling 
$f(\lambda)\le 0$ for each theory. 
In a setup of string theory 
($D=10, d=4$), $\lambda$ 
should satisfy 
\Eq{
\lambda\le -0.875001\,,~~~~
0.587274\le \lambda \le 19.0671\,.
}
\begin{figure}[h]
 \begin{center}
\includegraphics[keepaspectratio, scale=0.5, angle=0]{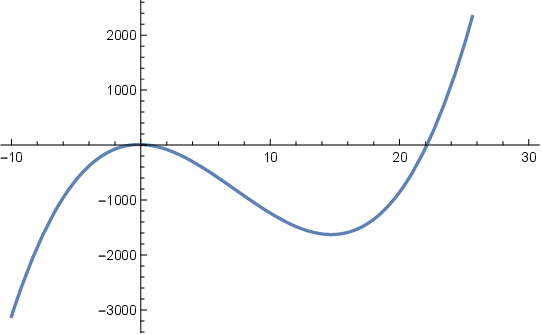}
\put(-105,90){$f(\lambda)$}
\put(0,55){$\lambda$}
\hskip 1cm
\includegraphics[keepaspectratio, scale=0.5, angle=0]{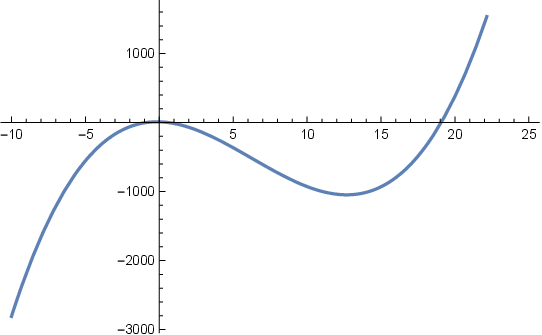}
\put(-105,90){$f(\lambda)$}
\put(0,60){$\lambda$}
\hskip 1cm
\includegraphics[keepaspectratio, scale=0.5, angle=0]{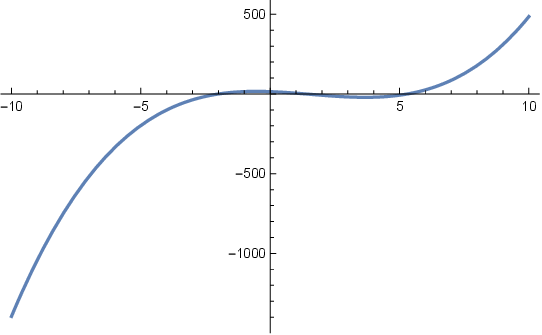}
\put(-77,90){$f(\lambda)$}
\put(0,65){$\lambda$}
\\
(a) \hskip 5cm (b) \hskip 5.5cm (c)~~~~~~
  \caption{\baselineskip 14pt 
The allowed region of the parameter $\lambda$ for $d=4$ 
satisfying $f(\lambda)\le 0$ is depicted in the case of 
$D=11$ (a), $D=10$ (b), and $D=5$ (c). One can recognize 
that the lower bound of $\lambda$ increases as the number of 
total dimension $D$ decreases.
}
  \label{fig:10}
 \end{center}
\end{figure}
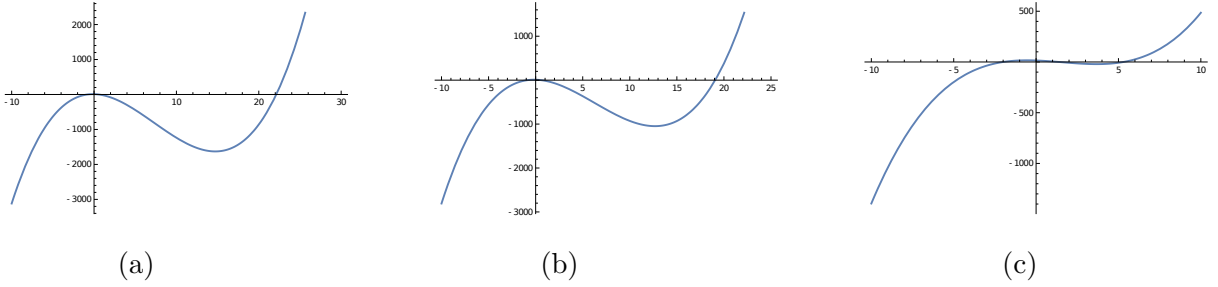
For the eleven-dimensional supergravity ($D=11, d=4$), 
we find 
\Eq{
-0.797148\ge \lambda\,,~~~~
0.54559\le \lambda \le 22.1224\,.
}
%
We also consider the five-dimensional supergravity, 
the the slope of the scalar potential $\lambda$ 
should satisfy 
\Eqr{
-1.99667\ge \lambda\,,~~~~~
1.38786\le \lambda\le 5.30366\,.
}
%

In the cases of exponential potentials that we investigated, an upper bound, 
$\lambda<\sqrt{3}$, was found from the minimal phenomenological requirements of a past epoch of radiation domination and acceleration today, while the simplest string theory realizations of the exponential potential in a flat universe have $\lambda>\sqrt{2}$ \cite{Agrawal:2018own, Akrami:2018ylq, Raveri:2018ddi,  Schoneberg:2023lun, Andriot:2024jsh}.
On the other hand, there is constraint of $\lambda<0.85$ (95\% CL) when using all currently available data such as BAO and Pantheon, and $\lambda<0.60$ (95\% CL) when adding a prior on $H_0$ from \cite{Schoneberg:2023lun}. However, surprisingly we find constraints to previous works \cite{Obied:2018sgi}, leading to a constraint of $\lambda>0.6$ in the string theory. Thus, there is a mild tension on the de Sitter swampland criterion, 


{\bf Previous treatments of the de Sitter conjecture have directly applied the results of \cite{Obied:2018sgi} to observational cosmology without adequate consideration of the underlying assumptions. While this approach remains valid in regimes where scalar field dynamics are negligible $\dot{\phi}=0$\,, substantial discrepancies arise when field dynamics play a significant role, as in quintessence scenarios. Our analysis demonstrates that proper inclusion of field kinetic terms yields parameter spaces consistent with four-dimensional observational cosmology. This approach represents a conceptually important advancement and constitutes a genuinely novel contribution, providing a rigorous analytical framework previously unavailable in the literature.}

\section{Summary and discussions}
  \label{sec:summary}

In this paper, we have discussed the de Sitter conjecture criteria when the time derivative of the scalar field cannot be neglected. When considering the quintessence model, this term is no longer  negligible. In such cases, using the original de Sitter conjecture criteria \cite{Obied:2018sgi} which is derived by completely ignoring the time derivative of the scalar field to discuss the validity of the quintessence model may no longer be entirely appropriate. 


We have focused on the quintessence model with an exponential potential of the scalar field and modified the criteria of the de Sitter conjecture by considering the new contributions emerging from the time derivative of the scalar field. We have found that, particularly when the higher-dimensional spacetime is 5 or 6 dimensions, the contributions generated by the time derivative terms cannot be ignored. If we consider quintessence models, which always have the potential dominating over the kinetic term, we will encounter the fine-tuning and coincidence problems as the cosmological constant, lacking the basin of attraction of tracker models. Hence, it is necessary for us to treat a contribution of the kinetic term of scalar field for a successful model. Moreover, our criteria of de Sitter conjecture differ in the allowed region of potential slopes compared to the original de Sitter conjecture. 

This offers the potential to resolve discrepancies between supernova and galaxy observations and four-dimensional dark energy models.
Of course, our model only deals with a gravity and scalar field system, effectively ignoring the presence of other matter fields and radiation. Therefore, we cannot definitively assert that our results will immediately bridge the gap between observations and theory. However, by discussing the novel contributions proposed in this paper within a complete dynamical system, we may take a step closer to resolving these issues. 


{\bf Regarding the accelerated expansion solutions in string theory, one correctly notes the inherent difficulty in realizing the current accelerating expansion of universe under strong energy conditions without incorporating scalar field dynamics. However, our analysis demonstrates that when the scalar field dynamics are properly included rather than neglected, the prospects for achieving cosmological acceleration are significantly enhanced. Our central argument is that in constructing four-dimensional accelerating cosmologies, accounting for scenarios where the kinetic contributions of fields such as quintessence remain non-negligible allows us to identify parameter regimes consistent with observational cosmology - a result that was not attained in \cite{Obied:2018sgi}.}

The dS conjecture \cite{Obied:2018sgi} cannot hold because of the problem with the Higgs potential that we discovered in \cite{Denef:2018etk}. 
This has not yet been resolved (see also \cite{Choi:2018rze, Han:2018yrk}).
Our new criteria also fall victim to the Higgs potential problem.

In our paper, the strong energy condition is assumed in the 
higher-dimensional theory. We assume that there is no Higgs potential
as described by \cite{Denef:2018etk} in the higher-dimensional theory. If there
is a Higgs potential in original higher-dimensional theory, we have to
consider the equation of motion of the higher-dimensional Higgs field
in the process of deriving equation (\ref{asw:new:Eq}) in our paper. However, if the Higgs potential of \cite{Denef:2018etk} appears "effectively" in the four-dimensional effective theory, the new
criteria will not solve the Higgs problem. Even if the de
Sitter criteria have a kinetic term of scalar field, the Higgs
potential has a much larger contribution than the quintessence
potential. The strong energy condition imposed in the paper will be
violated at least at the extreme value of the Higgs potential.
Even in this case, it would be ideal to choose the
ansatz of the metric well and consider a compactification in which the
Higgs and quintessence potentials are described by \cite{Denef:2018etk}.
Furthermore, if we impose a strong energy condition in the 
four-dimensional effective theory, it will give a constraint on
$\lambda$.

\begin{acknowledgments}
We thank Arthur Hebecker for valuable comments. The work of MN is supported in part by 
 JSPS KAKENHI [Grants No. JP22H01221 and JP23K22492] and the WPI program ``Sustainability with Knotted Chiral Meta Matter (WPI-SKCM$^2$)'' at Hiroshima University. 
 The work of K. U. is supported by Grants-in-Aid from the Scientific 
Research Fund of the Japan Society for the Promotion of 
Science, under Contract No. 16K05364 and by the Grant ``Fujyukai'' 
from Iwanami Shoten, Publishers. 

\end{acknowledgments}

\appendix




\end{document}